\documentclass[useAMS]{mn2e}

\usepackage{graphicx}

\def\be{\begin{equation}}
\def\ee{\end{equation}}
\def\ba{\begin{eqnarray}}
\def\ea{\end{eqnarray}}
\def\de{\partial}
\def\msun{M_\odot}

\def\ltsima{$\; \buildrel < \over \sim \;$}
\def\simlt{\lower.5ex\hbox{\ltsima}}
\def\gtsima{$\; \buildrel > \over \sim \;$}
\def\simgt{\lower.5ex\hbox{\gtsima}}

\title[HD molecules in merging haloes]{Formation of HD molecules in merging dark matter haloes}
\author[Yu. A. Shchekinov and E. O. Vasiliev]
       {Yu. A. Shchekinov$^1$\thanks{E-mail:yus@phys.rsu.ru}
and E. O. Vasiliev$^2$\thanks{E-mail:eugstar@mail.ru} \\
$^1$Department of Physics, University of Rostov,
$^2$Institute of Physics, University of Rostov\\
Sorge St. 5, Rostov-on-Don, 344090 Russia\\
}
\begin{document}
\date{Accepted 3004 December 15.
      Received 2004 December 14;
      in original form 2004 December 31}
\pagerange{\pageref{firstpage}--\pageref{lastpage}}
\pubyear{3004}
\maketitle

\label{firstpage}

\begin{abstract}
HD molecules can be an important cooling agent of the primordial gas
behind the shock waves originated through mergings of the dark matter haloes
at epochs when first luminous objects were to form. We study the necessary
conditions for the HD cooling to switch on in the low temperature range
$T<200$ K. We show that these conditions are fulfiled in merging haloes with
the total (dark matter and baryon) mass in excess of $M_{\rm cr}\sim
10^7[(1+z)/20]^{-2}\msun$. Haloes with masses $M>M_{\rm cr}$ may be the
sites of low-mass star formation.
\end{abstract}

\begin{keywords}
early Universe -- cosmology:theory -- shock waves.
\end{keywords}

\section{Introduction}

Cooling of primordial gas is determined by molecular hydrogen,
which forms in an expanding universe after recombination (Lepp \& Shull 1983,
Puy et al 1993, Palla, Galli \& Silk 1995, Galli \& Palla 1998,
Stancil, Lepp \& Dalgarno 1998). Inside the first virialized dark matter
haloes the fraction of H$_2$ can reach $10^{-3}$, which is able to cool gas
to 200~K. At lower temperatures energy losses in roto-vibrational H$_2$ lines
become insufficient to cool gas further. On the other hand, owing to a low
rotational energy and a large dipole moment HD molecules might be an efficient
coolant at $T<200$~K.  However, the abundance of HD is fairly sensitive to
thermal history of the gas, so that the question of whether HD dominates
radiative cooling at low temperatures or its contribution is negligible depends
on physical conditions along the evolutionary path. This circumstance is
reflected in conradictory conclusions about the role of HD in thermodynamics
of primordial gas: for instance, Lepp \& Shull (1983), Bromm, Coppi \& Larson
(2002) point out that HD cooling cannot dominate in primordial gas when the
first objects form, while Bougleux \& Galli (1997), Puy \& Signore (1997, 1998),
and more recently Uehara \& Inutsuka (2000), Flower (2002), Nakamura \& Umemura (2002),
Flower \& Pineau des Forets (2003) and Machida et al (2005) find that HD
molecules form in significant amount and can play a dominant role in thermal
evolution of pregalactic gas.

Transformation of H$_2$ molecules into their isotop analog HD is energetically
favoured because of higher binding energy of HD molecules. In equilibrium
(Solomon \& Woolf 1973, Varshalovich \& Khersonskii 1976)

\be
{n({\rm HD})\over n({\rm H_2})}=2{n({\rm D})\over n({H})}e^{465/T},
\ee
where factor 2 stems from the difference between the chemical constants
of HD and H$_2$ molecules: $\chi_{_{\rm HD}}-\chi_{_{\rm H_2}}=\ln 2$
(Landau \& Lifshitz 1969).
It is seen, that at $T\simlt 150$ K radiative HD cooling can enhance chemical
fractionation of deuterated molecules progressively: a small decrease in
temperature results in an increase of $n({\rm HD})/n({\rm H}_2)$, what in
turn increases radiative cooling in HD lines (Shchekinov, 1986).
In realistic conditions thermal equilibrium can be reached only in a characteristic
time depending on a set of chemical reactions which control the
transformation H$_2\Longleftrightarrow$HD. In primordial gas H$_2$ converges into
HD most efficiently in the reaction H$_2$+D$^+\rightarrow$ HD+ H$^+$
(Palla et al. 1995); other channels are unimportant on the Hubble time.
The background H$_2$ abundance and the frozen-out post-recombination fractional
ionization $x\sim 10^{-4}$ are sufficient to support deuterated H$_2$ only at
the level $n({\rm HD})\simlt 10^{-3}n({\rm D})$. Thus, a necessary condition
for the H$_2\rightarrow$ HD convergence to be efficient is that both H$_2$ and
electrons are abundant. In a cosmological pregalactic substrate such
conditions are most naturally fulfiled behind shock waves associated with
formation and virialization of dark matter haloes (Tegmark et al. 1997,
Barkana \& Loeb 2001). When formed in sufficient amount HD molecules can
cool gas down to the temperature of the CMB radiation -- the lowest possible
temperature, and thus provide conditions for formation of low-mass stars.
It is clear that this can occur with shock waves whose parameters (for instance,
postshock temperature, total mass involved, and so on) have reached some critical
values.

In the hierarchical scenario dark matter haloes of larger mass form through
mergings of smaller objects -- minihaloes (see for review Barkana \& Loeb 2001,
Ciardi \& Ferrara 2004). In these conditions the baryonic component of dark
haloes is compressed by shock waves. Apparently, all subsequent evolution of
baryons in dark haloes and their condensation into stars is determined by these
shock waves. Futhermore, one can consider that the virial state of dark matter
haloes and their baryons is reached through formation of shocks. It is clear,
that the physical state of the baryons processed by shocks depends on many
factors, such as geometry of a shock front, duration of a compressing flow etc.
However, basics characteristic features can be understood in the framework of a
head-on collision of two clouds of equal masses (Suchkov, Shchekinov \& Edelman
1983, Shapiro \& Kang 1987). In the context of the hierarchical scenario a
possible role of shock waves in enhancing H$_2$ cooling and stimulation of
primordial star formation was studied by Yamada \& Nishi (1998), and more
recently by Cen (2005). Radiative cooling in H$_2$ lines was indeed found
extremly efficient in decreasing the postshock temperature to $\sim 100$ K. In
spite of the fact that HD molecules are expected to form at this temperature,
HD chemistry was not taken into consideration in this analysis. The main goal
of our study are the conditions when HD molecules can be the dominant cooling
agent in a pregalactic gas.

In Section 2 we describe a simple thermo-chemical model of a shocked gas;
in Section 3 the conditions for HD molecules to be the dominant cooling agent
are analyzed; in Section 4 we discuss qualitatively a possibility of
a cold compressed layer to fragment; the discussion and summary of the results
are given in Section 5.

Throughout the paper we assume a $\Lambda$CDM cosmology with the parameters
$(\Omega_0, \Omega_{\Lambda}, \Omega_m, \Omega_b, h ) =
(1.0,\ 0.71,\ 0.29,\ 0.047,\ 0.72 )$ as inferred from the Wilkinson Microwave
anisotropy Probe (WMAP), and deuterium abundance $2.62\times 10^{-5}$
(Spergel et al. 2003).

\section{Thermo-chemistry of postshock gas}

For simplicity, we assume a head-on collision of two identical minihaloes,
which are virialized at redshift $z$ and have density of matter
$18\pi^2\Omega_m \rho_c (1+z)^3$, $\rho_c$ being the critical density; the
collision is assumed to take place at the same redshift $z$.
The minihaloes move with the relative velocity $v_c = \sqrt{3}\sigma$,
where $\sigma$ is a 1D velocity dispersion of the larger halo
\begin{equation}
\label{vel}
  \sigma^2 = {GM\over 2R} = GM^{2/3}(3\pi^3 \Omega_m \rho_c)^{1/3}(1+z).
\end{equation}
In the center of mass a discontinuity forms at the symmetry plane, and two shocks
propagate outwards. The time for the shock to pass through the entire minihalo
is $t_c = 3R/2v_c$. As it was pointed out by Gilden (1984) for supersonic
collisions the rarefaction time in transverse direction is greater than the
collision time, which means that a 1D scheme describes the overal picture
qualitatively correct. We assume the post-shock flow to be isobaric (Suchkov
et al. 1983, Shapiro \& Kang 1987, Anninos \& Norman 1996). The initial
post-shock temperature is
\begin{equation}
\label{tshock}
T_0 = {1 \over 3}{m_p v_c^2 \over k},
\end{equation}
where the post-shock gas is assumed to be isothermal, i.e. temperatures
of the electrons, ions and neutrals are kept always equal
$T_e=T_i=T_n$.

Chemical kinetics and thermal evolution of the gas are described by the
following set of equations
\begin{equation}
\label{spe}
\dot x_i = F_i({\bf x},T,n) - D_i({\bf x},T,n),
\end{equation}
\begin{equation}
\label{temp}
\dot T={2\over 3k}\sum\limits_i(\Gamma_i({\bf x},T,n) - \Lambda_i({\bf x},T,n))+{2\over 3}
{T\over n}\dot n,
\end{equation}
where $k$ is the Boltzmann constant, $x_i$ is the fraction of $i$-th species,
$F_i({\bf x},T,n)$, $D_i({\bf x},T,n)$ are respectively the formation and
destruction rates, $\Gamma_i({\bf x},T,n)$, $\Lambda_i({\bf x},T,n))$ are the
heating and cooling rates; ${\bf x}$ includes 14 species: H, H$^+$, H$^-$, He,
He$^+$, He$^{++}$, H$_2$, H$_2^+$, D, D$^+$, D$^-$, HD, HD$^+$, $e$.
In the energy equation (\ref{temp})
all relevant cooling processes (such as collisional excitation and
ionization, Compton cooling, recombination and bremsstrahlung radiation,
cooling by HI, H$_2$ and HD line emission) are accounted for. For the cooling
functions of H$_2$ and HD we adopted the expressions given by Hollenbach \&
McKee (1979), and Flower et al. (2000), respectively; note, that the
Hollenbach \& McKee H$_2$ cooling function practically coincides with the
one calculated by le Bourlot, Pineau des Forets \& Flower (1999). The most recent
calculations of the HD cooling function by  Lipovka, N\'u\~ nez-L\'opez, \&
Avila-Reese (2005) show that in the temperature range of interest ($T<10^3$ K)
it coincides with that found by Flower et al. (2000). At higher
temperatures, where Lipovka et al. (2005) predict an order of magnitude
enhanced cooling from HD, the abundance of HD molecules is too low to
contribute significantly. 
The other cooling functions connected with excitation and ionization of
H and He are taken from Cen (1992).
In general, radiation from a hot shocked gas can heat the gas, however,
for the shock waves with the velocity $\simlt 60$ km s$^{-1}$ (Shull \& McKee
1979, Shull \& Silk 1979) its contribution to ionization and heating is
unimportant, therefore in our calculations we assume $\Gamma_i({\bf x},T,n)=0$.
The chemical reaction rates are taken from
Galli \& Palla (1998). The thermo-chemical equations (\ref{spe}), ({\ref{temp})
are solved in one collisional time $t_c$; $t_c$ is much shorter than
the age of the universe at any redshift $z$.

Fig.1 shows an example of physical characteristics of the postshock gas versus
temperature in the collision of two minihaloes with the total masses
$M = 10^7~\msun$ each at $z=20$. The fractional ionization $x$ and H$_2$
abundance behind the shock are taken equal to their background values
$x=10^{-4}$ and $f({\rm H}_2)=10^{-5}$, $f({\rm HD})=10^{-9}$.
Immediately behind the shock concentration of H$_2$ increases very rapidly and
reaches its maximum value $\sim 2\times 10^{-3}$ promoting cooling of gas to
$T\sim 200$ K. If H$_2$ cooling decreases the gas temperature below
$T=150-160$ K the abundance of HD molecules grows fast and HD cooling becomes
dominant. It sets a runaway cooling on. At $T<100$ K almost all the deuterium
is converted into molecular form, and in one collisional time  gas cools
down to $T\simeq 60$ K.

\begin{figure*}
\vspace{24pt}
\includegraphics[width=85mm]{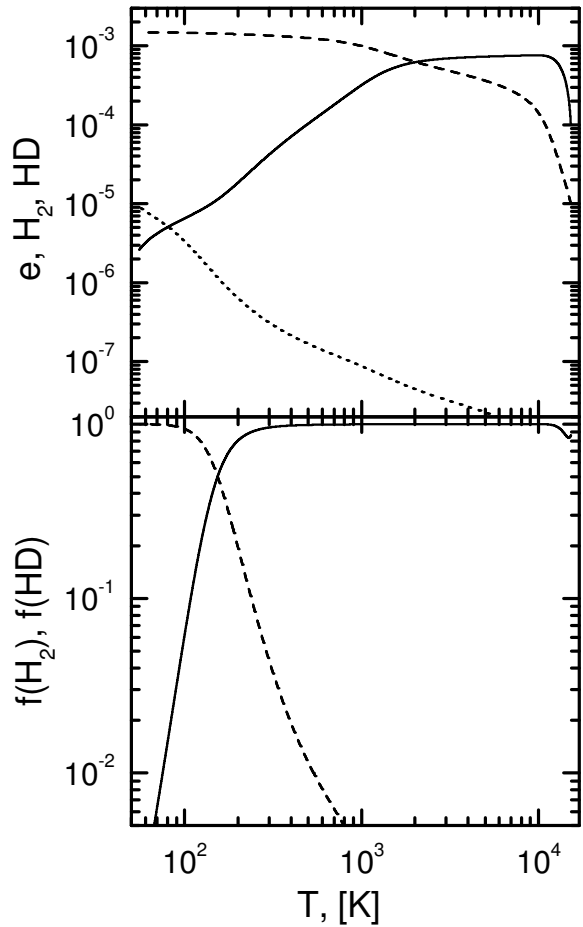}
\caption{Thermo-chemical evolutionary state vs temperature of a
               postshock gas parcell in the collision of minihaloes with
               the mass $M = 10^7~\msun$ each at $z=20$:
               in the upper panel abundances of the species: electrons
               (solid), H$_2$ (dashed) and HD (dotted) lines -- are shown;
               the lower panel shows the relative contribution of H$_2$ (solid)
               and HD (dashed) molecules to the total cooling rate.
               }
         \label{Fig1}
\end{figure*}

Haiman, Rees, \& Loeb (1997) showed that the very first stars emit a
large amount of photons in Lyman-Werner band (LW), which produces a hostile
UV background for HD molecules. The characteristic time of photodissociation
of HD can be estimated as
$t_{\rm diss} = 10^4/J_{\rm LW}$~yr, where $J_{\rm LW}$ is
the LW flux in units of
$10^{-21}$~ergs~s$^{-1}$~cm$^{-2}$~Hz$^{-1}$~sr$^{-1}$
(Le Petit, Roueff, \& Le Bourlot, 2002, Johnson \& Bromm, 2005). One
can assume that the LW flux is by an order of magnitude equal to the ionizig
flux from the first stars $J_{\rm LW} \simeq J_{21}\sim \exp[-(z-5)]$ in the
range $5\leq z \leq 20$ (Ciardi et al., 2000).
On the other hand, the characteristic formation time
of HD is $t_{\rm form} \sim 3\times 10^6$~yr at $z=10-20$.
Thus, practically in the whole range of redshifts considered here
photodissociation from the very first stars can be neglected:  
$t_{\rm diss}$ becomes comparable to $t_{\rm form}$ only at $z\simlt 10$.
At lower $z$ self-shielding seems to come into play. Indeed, for the masses
of merging haloes $M>10^6\msun$ the characteristic column density of HD
molecules in the shocked layer at final stages can be as high as
$\sim 3\times 10^{15}$ cm$^{-2}$, which is factor of 3 greater than the
column density needed to provide optical depth in LW band of the order 
of one $\sim 10^{15}$ cm$^{-2}$ (Le Petit et al., 2002).

\section{HD behind shock waves}

Haloes of small masses ($M<5\times 10^6\msun$) merge with slow relative
velocities, and the post-shock temperature is so low that fractional
ionization remains equal to its post-recombination value $x\sim 10^{-4}$.
As a result, $(x,T)$ diagrams connecting $x(t)$ and $T(t)$ along evolutionary
paths very weakly depend on halo masses, as seen from Fig. 2.
Mergings of more massive haloes produce shocks with temperature
$T\simgt 10^4$ K, so that fractional ionization can be as high as
$x\sim 10^{-2}$. Immediately behind the shock $x$ starts growing from its
pre-shock value to the corresponding equilibrium. However, as the
characteristic collisional ionization time is longer than the cooling time,
gradual increase of fractional ionization is accompanied by cooling, which in
turn weakens the rate of collisional ionization. As a result, fractional
ionization remains always lower that the equilibrium value determined
by collisional ionization and radiative recombination. At temperature
$T\simeq 10^4$ K $x$ reaches the maximum which depends on the postshock
temperature, and then goes down as shown in Fig. 2. This behavior restricts
the efficiency of molecule formation: as seen from Fig. 1 H$_2$ saturates at
$T\simlt 10^3$ K when fractional ionization begins to decrease sharply. HD
instead monotonously grows until practically all deuterium is converted into
molecular form. The abundance of H$_2$ molecules at the end of baryon
compression increases with the mass of the merging haloes,
and the minimum temperature reached at this stage becomes lower. When it
reaches $T\simlt 150$ K a rapid fractionation of HD molecules sets on. For
larger halo masses HD cooling dominates and cools baryons down to the
temperature of the CMB radiation. Fig. 3 connects variations of temperature
and  molecular concentrations (H$_2$ and HD) along the evolutionary paths
for different masses of the haloes merging at $z=20$. Their contribution to
the total cooling is shown in Fig. 4 where open circles show the ending stage
after one collision time $t_c$: while immediately after crossing the shock
front the contribution from HD (left curve) is negligible,
it increases in the course
of gas cooling, and the transition to the regime with
a predominance of HD cooling is clearly seen at $M\simeq 10^7\msun$.
The critical mass corresponding to the onset of a predominantly HD cooling is
$M_{\rm cr}^{\rm HD}\sim 8\times 10^6[(1+z)/20]^{-2}\msun$ as shown in
Fig. 5. This value is factor of 5 higher than the minimum mass of the first
objects found by Tegmark et al. (1997). One may conclude therefore, that at
the lower end of masses the primeval gravitationally bound baryonic condensations
stop cooling at relatively high temperature $T\sim 200$ K due to H$_2$ radiation,
and HD molecules do not form in sufficient amount to cool the baryons further.
However, soon later, at $z\simeq 18$ more massive haloes can form and merge
with a clear predominance of HD cooling, and in principle with lower masses of
possible stars. Note, that the critical mass $M_{\rm cr}^{\rm HD}$ is fairly
unsensitive to the choice of cooling function. In particular, for the cooling
function of Galli \& Palla (1998), who included the collisional rates from
Martin, Schwarz \& Mandy (1996) and Forrey et al. (1997), which in the
temperature range $T\simlt 10^4$ K is approximately factor of 3 higher than
H$_2$-cooling function of Hollenbach \& McKee in our calculations,
$M_{\rm cr}^{\rm HD}$ increases only by 10--15\% in comparison with that shown
in Fig. 5.

\begin{figure*}
\vspace{24pt}
\includegraphics[width=85mm]{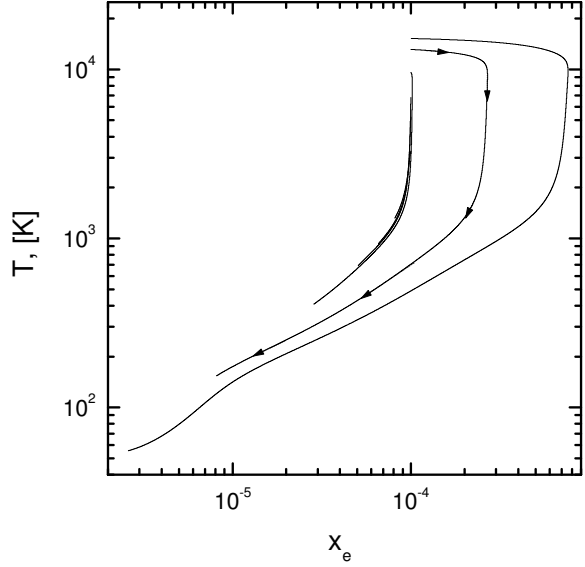}
  \caption{The evolutionary paths connecting fractional ionization and
               gas temperature behind shocks in collisions of
               minihaloes of masses $10^6$, $2\times 10^6$, $3\times 10^6$,
               $5\times 10^6$, $8\times 10^6$, $10^7$ (from left to right) at
               $z=20$. The arrows show the direction of evolution of a
               gas parcel starting from its state immediately behind the shock;
               paths for the first four masses practically coinside.
               }
         \label{Fig2}
\end{figure*}

\begin{figure*}
  \vspace{24pt}
\includegraphics[width=155mm]{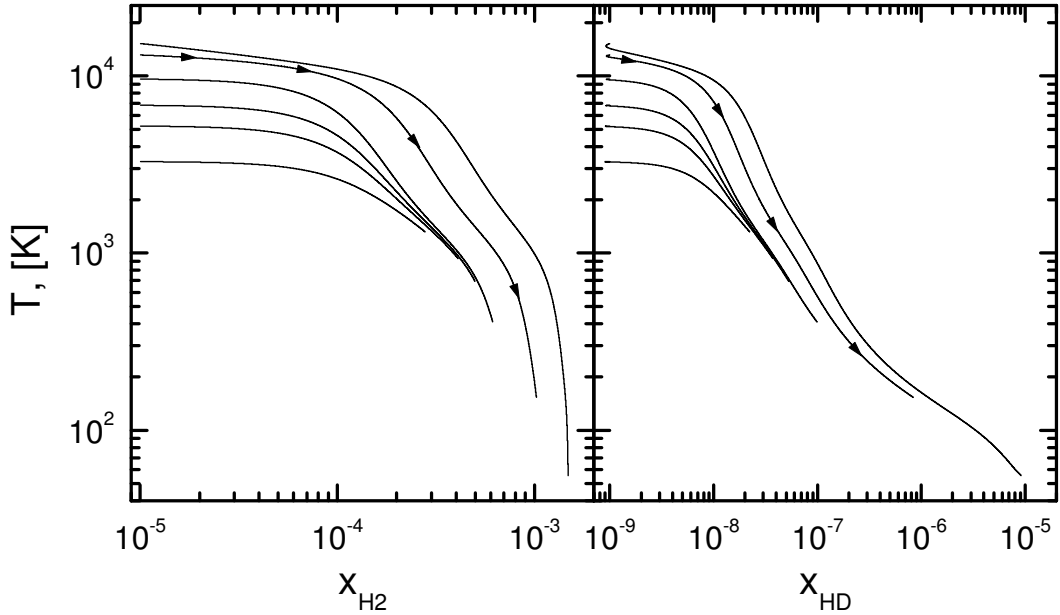}

     \caption{The evolutionary paths connecting variations of
              temperature and concentrations of H$_2$ and HD molecules for the same
              masses as in Fig. 2 (from left to right).
               }
         \label{Fig3}
\end{figure*}

\begin{figure*}
  \vspace{24pt}
\includegraphics[width=85mm]{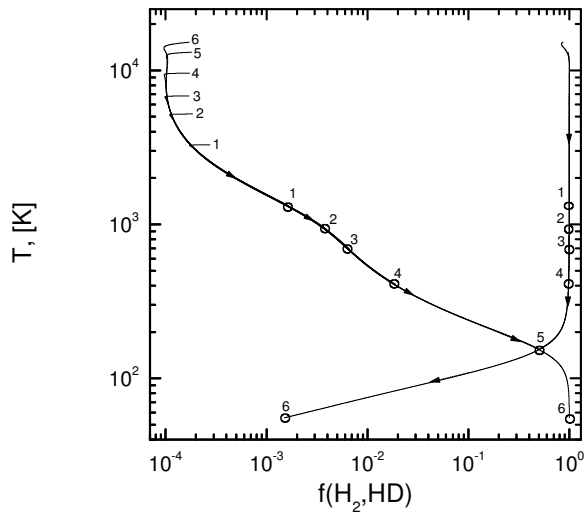}

     \caption{The relative contribution of H$_2$ (right curve) and
     HD (left curve) into the total cooling
     for the same set of models shown in Fig. 2: increasing numbers correspond
     to the increasing masses in Fig. 2; open circles mark the state after one
     collision time $t_c$. }
         \label{Fig4}
\end{figure*}

\begin{figure*}
  \vspace{24pt}
\includegraphics[width=85mm]{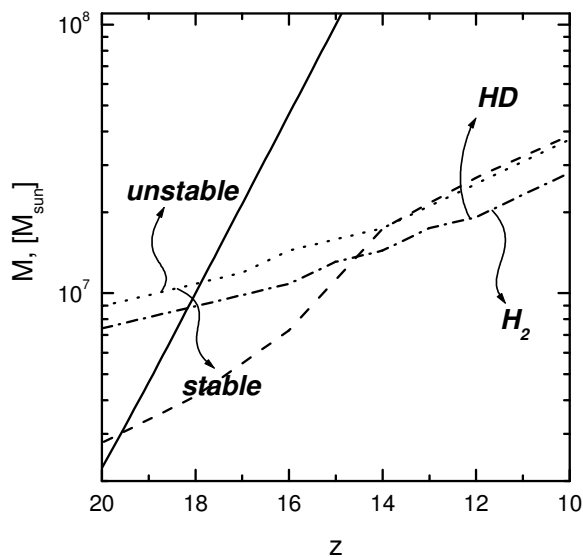}

      \caption{The mass-redshift diagram delimitating interval of halo masses,
               where HD molecules dominate in cooling at low temperatures
               (this is the range above the dash-dotted line), and halo masses
               with the postshock baryon layer being gravitationally unstable
               and subjected to further fragmentation (these masses above the
               dotted line). The thick solid line shows the 3$\sigma$
               fluctuations in the $\Lambda$CDM cosmology; the dashed line
               shows the minimum mass limit found by Tegmark et al (1997).
}
         \label{Fig5}
\end{figure*}

\begin{figure*}
  \vspace*{24pt}
\includegraphics[width=85mm]{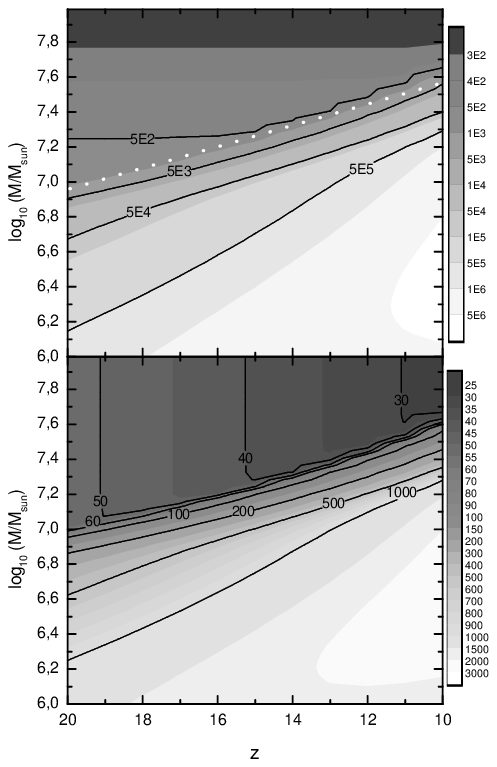}
      \caption{Jeans mass (upper plate) and temperature (lower plate)
               reached at one collision time in the postshocked gas
               as a function of the halo mass and the redshift when the haloes
               merge. White dotted line marks masses of the haloes whose
               compressed layers are gravitationally unstable. Scale bars in
               the right are for Jeans mass (upper) in solar units, and gas
               temperature (lower) in K.
               }
         \label{Fig6}
\end{figure*}

\section{Fragmentation of the cold gas layer}

When cooled the shocked gas layer can fragment due to gravitational
instability (Stone 1970, Elmegreen \& Elmegreen 1978, Gilden 1984). For a
layer formed in a collision of two clouds the necessary conditions for
instability imply that {\it i)} the characteristic growth time is shorter
than the collision time, and {\it ii)} the critical wavelength is shorter
than the initial size of the clouds (Gilden 1984). The dotted line in Fig. 5
depicts the critical mass $M_{\rm cr}^{\rm G}$ of a merging halo
gravitationally unstable in Gilden sense: merging haloes with individual
masses above this line are unstable againts fragmentation;
$M_{\rm cr}^{\rm G}$ is approximated as
$M_{\rm cr}^{\rm G}\simeq 10^7[(1+z)/20]^{-2}\msun$. It is readily seen
that $M_{\rm cr}^{\rm G}$ is practically coincident with the mass
corresponding to the transition to a predominance of HD cooling
$M_{\rm cr}^{\rm HD}$. The explanation of this coincidence can be found in the fact
that the shock compressed layers become unstable in Gilden sense only when
HD molecules cool the compressed gas below $T\simeq 150$ K (Vasiliev \&
Shchekinov 2005). It is obvious therefore
that all masses $M>M_{\rm cr}^{\rm G}$ at the latest stages are dominated by HD
cooling, and as a consequence
the fragments compressed by mutual action of ram
pressure and gravitation are able to cool further down to $T<100$ K. In the
layer formed in a collision of two haloes with masses $M=10^7\msun$ the
corresponding Jeans mass is $M_J=10^3\msun$, which is less than 1\% of the
baryonic mass involved into collision. Fig. 6 shows the dependence of Jeans
mass in unstable layers on the mass of merging haloes and the redshift $z$
when the haloes merge: thick dotted white line delimits the haloes whose
compressed layers are gravitationally unstable -- all masses above this line
are unstable. The thin solid lines are the iso-Jeans masses in compressed
layers. When temperature in the compressed layer goes to the CMB value the
Jeans mass becomes independent on redshift. Clarke \& Bromm (2003) have
estimated a lower limit of the masses of fragments formed in similar
conditions, with baryons confined by ram pressure from shocks and gravitation.
However, for the primordial gas they assumed the lower temperature
limit $T=200$ K determined by H$_2$
cooling. As a result, the masses of the fragments are systematically higher
compared to our estimates. In order to obtain smaller values of the masses
they had to assume that the compressed gas has cooled down to the CMB
temperature due to cooling provided by CO molecules
only after the gas is pre-enriched with metals.

The lower panel of Fig. 6 shows the gas temperature at the latest stages of
gas compression, i.e. at $t=t_c$, for mergings of haloes with masses $M$
at redshift $z$. It is clearly seen that the compressed gas can cool down
to the lowest temperature $T_{\rm CMB}(z)$ in haloes with masses
$M>M_{\rm cr}^{\rm HD}$ as their cooling is dominated by HD molecules.
Note, that this result does not contradict the conclusions drawn from
numerical simulations by Abel et al. (2000, 2002) and Bromm et al. (2002),
because these simulations have been done for halo masses close to the minimum
mass expected for first objects.
\begin{figure*}
  \vspace{24pt}

\includegraphics[width=85mm]{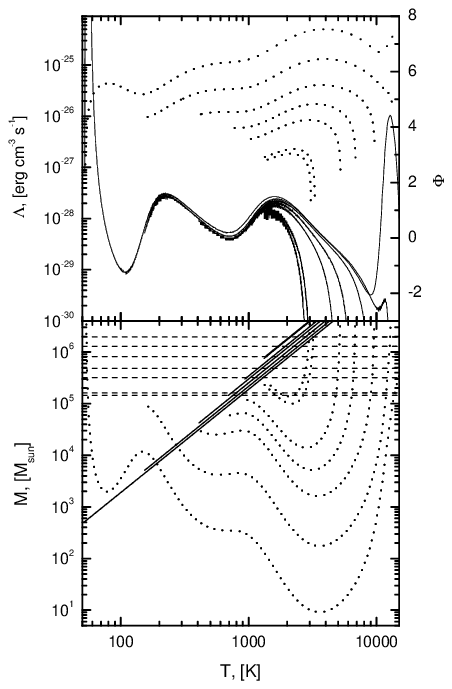}
      \caption{Upper panel: The temperature dependence of the effective cooling function
               $\Lambda(T)$ (dashed line), and its logarithmic
               derivative $\Phi=(\de\ln \Lambda/\de\ln T)_n$ (solid line). In a non-steady
               cooling medium temperature interval with $\Phi<2$ is thermally
               unstable. Initial density value is 1.5~cm$^{-3}$.
               Lower panel:
               Characteristic mass of condensations formed through thermal
               instability is shown by dotted lines. Solid lines show Jeans
               mass in the shocked layer.
               The solid and dotted lines on both panels are for halo masses
               $M_{\rm h}=9\times 10^5, 10^6, 2\times 10^6, 3\times 10^6,
               5\times 10^6, 8\times 10^6, 1.2\times 10^7 \msun $
               from top to bottom. Each curve is shown in the temperature
               interval from the initial value $T_0$ to the lowest value
               reached at the end of compression $t=t_c$ for a given halo
               mass. Horizontal short-dashed
               lines show the baryon mass of the same
               set of halo masses in a reversed order -- from bottom to top. }
         \label{Fig7}
\end{figure*}

Fragmentation of a compressed gas layer can be initiated by thermal instability
at the stages when H$_2$ and HD begin to dominate radiative cooling. In Fig. 7
(upper panel) temperature dependence of the effective ccoling function
$\Lambda(T)$ and its logarithmic derivative $\Phi=(\de\ln \Lambda/\de\ln T)_n$
are shown in the temperature interval $T<10^4$ K, where HI radiative energy
losses become unimportant and molecular cooling sets on. The curves
corresponding to different masses are plotted in temperature interval from the
initial value immediately behind the shock to the lowest value at $t=t_c$.
In a non-steady cooling medium without external heating the condition for
the isobaric mode to be unstable reads as $\Phi<2$ (e.g., Shchekinov 1978).
It is evident that this condition is fulfiled at $T\simlt 9000$ K due to
cooling of H$_2$ molecules, and in the lower temperature end at $T\simlt 100$ K
where HD cooling dominates. The characteristic length of thermally unstable
perturbations is $\lambda_{\rm R}\sim c_{\rm s} \tau_{\rm R}$, where

\be
\tau_{\rm R}\sim {3\over 2}{kT\over \Lambda },
\ee
is the characteristic radiative cooling time, $c_{\rm s}$ is the sound speed.
The corresponding masses
$M_{\rm R}\sim \rho_{\rm R}(\lambda_{\rm R}/2)^3$, estimated for the
parameters of a shocked gas, are shown in the lower panel of Fig. 7; here
$\rho_{\rm R}$ is the baryon density at the stage when thermal instability
begins. For the haloes of large masses the mass of fragments
which can form through thermal instability is lower than the total mass of baryons
and Jeans mass practically in the whole temperature range of compressed
gas layers. This means that in these conditions thermal instability
stimulates subsequent fragmentation of the compressed gas. The minimum mass
of fragments expected from thermal instability
$M_{\rm R}\simeq 10^{26}M_{\rm h}^{-3.5}\msun$ can be as small as
$M_{\rm R}<10\msun$ for halo masses $M_{\rm h}>1.2\times 10^7\msun$. However,
thermal instability does not work in the haloes of smaller masses when
$M_{\rm R}$ exceeds their baryon mass. As it seen from Fig. 7 this occurs at
$M_{\rm h}\leq 9\times 10^5\msun$. The clouds formed
under thermal instability are pressure supported, but not necessarily
gravitationally bound -- it is seen from Fig. 6 that thermal instability
forms gravitationally bound clouds of stellar masses only in mergings of
massive haloes $M\simgt (1-3)\times 10^7\msun$. Therefore not always thermal
instability can give rise directly to star formation, however collisions of
clouds formed through it can be a stimulating factor.

The compressed layers gravitationally unstable against fragmentation,
$M>M_{cr}^{\rm G}$, lie in the halo mass range where HD cooling can keep gas
isothermal with temperature close to $T_{\rm CMB}$. Therefore, once the
fragments are formed, they can afterwards either collapse homogeneously or
brake further on smaller masses as prescribed by the hierarchical
fragmentation scenario (Hoyle 1953). The possibility of the
fragmentation cascading down to lower scales still remains under debate
(see, e.g., discussion in Coppi et al. 2001, Glover 2005).
An essential condition for such a cascading process
is a pressure-free contraction regime. In practice, gravitational
collapse turns out to settle onto a slow quasi-hydrostatic regime, which
results in an efficient subsonic damping of perturbations (Abel et al. 2002,
Bromm et al. 2002), and therefore in a strong suppress of further
fragmentation.
The mass of dense clumps which can be reached in the hierarchical fragmentation
scenario gives, however, an estimate of the lower mass limit of protostellar
condensations expected at given physical conditions --
the opacity limit for fragmentation (Low \& Lynden-Bell 1976, Rees 1976).
This mass is determined by a maximum density when the forming fragments become
optically thick in the lines providing dominant radiative cooling.
When HD molecules are
the main cooling agent, a hierarchical fragmentation becomes
optically thick in HD rotational lines, i.e the optical depth of
a fragment at line centre becomes $\tau_{\rm HD}=1$,} at
densities $\sim 10^9-10^{10}$ cm$^{-3}$. Even though at these densities most
of hydrogen converts to molecular form in three-body collisions (Palla et al.
1983) its contribution to the total cooling remains negligible because of low
temperatures. The corresponding Jeans mass at this stage is
$M_J\sim 30T_{\rm CMB}^{3/2}n^{-1/2}\msun\sim 10^{-3}(1+z)^{3/2}\msun$
(Vasiliev \& Shchekinov 2005). For $z=10-20$ this gives a relatively low mass
limit: $M_{\rm J}\sim (0.03-0.1)\msun$. A similar estimate for the mass of
fragments formed behind the shock waves with velocities
$\geq 300$ km s$^{-1}$ from supernovae explosions was obtained by Uehara \&
Inutsuka (2000). One should stress though, that the condition
$\tau_{\rm HD}=1$ implies that photons from HD rotational lines experience
in average only one scattering, which seems insufficient to lock radiation
inside a contracting cloud and stop further decrease of Jeans mass. From
this point of view, the condition $\tau_{\rm HD}=3-5$ equivalent to
about 10 scatterings of photons (Ivanov, 1973), gives apparently more accurate
estimate of the opacity mass limit: the corresponding Jeans mass is
$M_{\rm J}\sim (0.01-0.03)\msun$. For comparison, the estimates of the
opacity mass limit
of the fragments determined by H$_2$ cooling vary from $3\msun$ to
about $10^2\msun$ (Nakamura \&
Umemura 1999) depending on the initial H$_2$ abundance.
Note, that the exact value of the opacity limit
depends on dynamical regime of a collapsing cloud: the optical depth of
a contracting cloud decreases with the velocity gradient as $|dv/dr|^{-1}$
(Sobolev, 1960). In numerical simulations
$|dv/dr|\simeq 2v_{\rm T}/R_{\rm J}$ at late stages of the contraction
(Abel et al. 2002), here $R_{\rm J}$ is the Jeans length. Abel et al. (2002)
stop their similation when optical depth at line centre is 10. With accounting
the factor 2 in the velocity gradient this corresponds to the effective
optical depth of 5 -- nearly at the beginning of the stage when radiation
becomes locked inside the contracting fragment.

\section{Conclusions}

In this paper we showed that

\begin{enumerate}

\item In the hierarchical scenario of galaxy formation mergings of
massive dark matter haloes, $M\simgt 8\times 10^6[(1+z)/20]^{-2}\msun$
develop dense baryon layers, where practically all the deuterium becomes
confined into HD molecules, which then provide a very efficient radiative
cooling;

\item Only slightly higher is the critical mass of the merging
haloes when the shocked baryon layers can be unstable against gravitational
fragmentation $M>10^7[(1+z)/20]^{-2}\msun$.

\item In these conditions the shocked baryons can cool down to the lowest
temperature $T\simeq T_{\rm CMB}=2.7(1+z)$ K;

\item The fragments formed through instability evolve then practically
isothermally with $T\sim T_{\rm CMB}$, and can in principle undergo further
sequencial fragmentation, which stops when the minimum mass
$\msun\sim 10^{-3}(1+z)^{3/2}\msun$ is reached.

\item This minimum mass is smaller than the one expected when H$_2$ molecules
control radiative cooling (Nakamura \& Umemura 2002,
Ciardi \& Ferrara 2005). This means that the first
stars can be less massive in galaxies with masses
$M>8\times 10^6[(1+z)/20]^{-2}\msun$, where HD cooling dominates.
\end{enumerate}


\bsp

\label{lastpage}

\end{document}